    \documentstyle[aps]{revtex}

    \def\half{{\textstyle{1\over2}}}
    
    \def\eigh{{\textstyle{1\over 8}}}

    \def\beq{\begin{equation}}
    \def\eeq{\end{equation}}
    \def\bi{\begin{itemize}}
    \def\ei{\end{itemize}}
    \def\beqar{\begin{eqnarray}}
    \def\eeqar{\end{eqnarray}}

    \newcommand{\Ee}{\mbox{$\cal E\;$}}

    \newcommand{\Pp}{\mbox{$\cal P\;$}}
    \newcommand{\bcP}{\mbox{\boldmath$\cal P$}}
    
    \def\vec#1{{\bf#1}}

    \newcommand{\rmd}{{\rm d\null}}
    \newcommand{\rmdd}[1]{\rmd^d#1\,}
    \def\boldnab{\mbox{\boldmath$\nabla$}}
    \def\boldr{{\bf r}}
    
    \def\pmb#1{\mbox{\boldmath$#1$}}

    \newcommand{\be}{\begin{equation}}
    \newcommand{\ee}{\end{equation}}
    \newcommand{\ea}{\end{eqnarray}}
    \newcommand{\ba}{\begin{eqnarray}}

\begin{document}
 \title{Hidden Symmetry of a Fluid Dynamical Model}
 \author{C. Neves$^2$ and C. Wotzasek$^1$}      
 \address{$(1)$Instituto de F\'\i sica\\
Universidade Federal do Rio de Janeiro\\
21945-970, Rio de Janeiro, Brazil\\
$(2)$ Departamento de F\'\i sica, ICE, Universidade Federal de Juiz de Fora,  
\\36036-330, Juiz de Fora, MG, Brasil,\\}
          \date{\today}     
    \maketitle

    \begin{abstract}
A connection between solutions of the relativistic d-brane system in (d+1) dimensions
with the solutions of a Galileo invariant fluid in d-dimensions is by now well established.
However, the physical nature of the light-cone gauge description of a relativistic membrane
changes after the reduction to the fluid dynamical model since the gauge symmetry is lost.
In this work we argue that the original gauge symmetry present in a relativistic d-brane system
can be recovered after the reduction process to a d-dimensional fluid model. To this end we
propose, without introducing Wess-Zumino fields, a gauge invariant theory of isentropic fluid dynamics and show that this symmetry corresponds
to the invariance under local translation of the velocity potential in the fluid dynamics picture.
We show that different but equivalent choices of the sympletic sector lead to distinct representations
of the embedded gauge algebra.
\end{abstract}

    \section{Introduction}    

    Some years ago, Bordemann and Hoppe\cite{BH} demonstrated that the relativistic
    theory of membranes are integrable systems by reducing the problem to a 2-dimensional
    fluid dynamics. In that paper the authors simplify the light-cone gauge description
    of a relativistic membrane\cite{JH} moving in a Minkowski space by changing the
    independent to dependent variables. This allowed them to find explicit solution to
    all constraints and to reduce the original system of field equations, with four
    functions, to a system with only two functions.  The dynamics is governed by a
    Hamiltonian reduced to a ${\it SO}(1,3)$ invariant (2+1)-dimensional theory of
    isentropic fluid dynamics, where the pressure is inversely proportional to the
    mass-density. Unfortunately, this procedure does not preserve the gauge symmetry
    exhibit initially by the relativistic theory of membranes, which can call into doubt
    the reduction procedure. Afterward, the study of the non-relativistic isentropic
    fluid mechanics model \cite{LL} has attracted much attention
    \cite{AJ,BJ,DB,JP,HH,NO}. This subject is of broader interested since it also offers
    connections with the hydrodynamical description of quantum mechanics \cite{EM,M},
    parton
    model \cite{AJ}, black-hole cosmology \cite{KMP}, hydrodynamics of superfluid systems
    \cite{AMJS}. Most of these investigations are dedicated to find the solutions of this
    Galileo invariant system in d-dimensions in connection with the solutions of the
    relativistic d-brane system in (d+1)-dimensions\cite{BJ,DB}, which is of direct
    interest to theoretical particle physics. In particular, this last point was
    responsible for the recent spate of interest in clarifying the presence of a hidden
    dynamical Poincare symmetry of this non-relativistic model realized by field
    dependent diffeomorphism: in terms of the canonical variables one can compute the
    Poisson algebra and reproduces the Poincare algebra for a system (membrane) in one
    dimension higher \cite{DB}.

    In this paper we argue that the U(1) gauge symmetry present on the light-cone
    description of a relativistic membrane can be preserved on the 2-dimensional theory
    of isentropic fluid dynamics obtained after the reduction via field-dependent change
    of variables\cite{BH} or dimensional reduction of a scalar relativistic field
    theory\cite{AJ}.
    We will ventilate that this reduced system indeed possess a U(1) gauge symmetry very
    much like the paradigmatic Maxwell theory, although it is a constrained system with a
    second-class description. Due to this second-class character, the presence of this
    local symmetry in this model was not suspected.
    Indeed, it has been pointed out in \cite{AMJS} that the classical hydrodynamics
    description of an isentropic fluid is invariant under global translation of the
    velocity potential. 
    Our goal in this paper is to recover the local U(1) gauge symmetry fixed by the
    reduction process, that was not disclosed in previous investigations, and
    consequently to find the dynamically equivalent gauge formulation of the noninvariant
    isentropic fluid by reverting the nature of the constraints.

    In this regard it is worth mentioning that there has been an intense research in second-class constraints conversion recently.
    The basic idea behind the conversion process is to identify a first-class subset of
    the constraints. This can be done by means of two distinct concepts. One path follows
    the proposal of Faddeev-Shatashvilli \cite{FS} of enlarging the phase-space with
    Wess-Zumino variables (WZ) \cite{BFFT,IJMP}. This has been shown to be possible even
    for nonlinear second-class constraints of arbitrary geometries\cite{NW,JNW}. Another
    path of conversion process is confined into the original phase-space from the outset
    \cite{Mitra,VT}. In this approach half of the constraints are seen as gauge-fixing
    conditions for the first-class subset. An appropriate projector operator is then
    constructed that maps all observable to the invariant sector of the theory. We shall
    follow this line in this work.

    We have organized this paper as follows. In Section 2, a review of the fluid dynamics
    model will be presented, paying particular attention to the complete set of
    symmetries. In section 3, the original second-class description of the fluid dynamics
    model will be reformulated as a first-class model, emphasizing that half of the set
    of constraints is assumed as gauge symmetry generator while the remaining constraints
    are considered as gauge-fixing terms. The corresponding gauge transformations are
    explicitly computed. We pay particular attention on the dependence of our choice of the original symplectic structure, which is deceptively trivial, leading to dramatic differences in the final results.
    The last section will be reserved to stress our conclusion and final discussions.
    In an appendix, we will sketch the gauge unfixing Hamiltonian method\cite{Mitra,VT}.

    \section{The fluid dynamics model}

    After the original reductive procedure of \cite{BH}, the problem of 2-dimensional
    flow\cite{BH,AJ} has been investigated intensively \cite{AJ,BJ,DB,JP,HH,NO}. An
    important contribution to this topic was given in \cite{AJ,BJ}, where two more
    conserved quantities and the associate symmetries were found. In this section we
    briefly review these results.

    There are many ways to derive the Lagrangian that determine the fluid dynamics of
    interest here. We start with the Schr\"odinger Lagrangian \cite{EM}, 

    \beq
    L_{\rm S} = \int d^d r \Bigl\{ i \psi^* \dot{\psi} -
    \half (\boldnab\psi^*)\cdot(\boldnab \psi) -
    \bar{V} (\psi^*\psi) \Bigr\},
    \label{0000}
    \eeq
    with $\bar{V}$ determining any nonlinear interaction. Introducing the representation
    in terms of mass density and velocity potential, as usual \cite{AMJS},
    \beq
    \psi = \rho^{1/2} e^{i\theta}
    \label{0010},
    \eeq
    into the Schr\"odinger Lagrangian, we obtain,

    \beq
    {\cal L} = \int d^d r\;\left(- \rho \dot \theta - \frac 12 \rho(\boldnab \theta)^2 -
    V(\rho)
    \right),
    \label{0020}
    \eeq
    with

    \beq
    V(\rho) = \bar{V} (\rho) + \eigh \frac{(\boldnab \rho)^2}{\rho}\, ,
    \label{0030}
    \eeq
    which is the hydrodynamical form of the Schr\"odinger theory\cite{EM,M}. Notice that
    there is a nontrivial {\it interaction}, even in the absence of $\bar{V}$.  This
    result was obtained from a gauge-fixed formulation of a membrane in Minkowski
    space\cite{BH}, through a field-dependent change of variables,  for the special case
    $d=2$ and a potential of a strength g given by,

    \beq
    V(\rho) = \frac{g}{\rho}.
    \label{0040}
    \eeq
    The same result was also obtained from a dimensional reduction of a local
    relativistic field theory \cite{AJ}.

    {}From this derivation, there is no doubt that the model described by Lagrangian
    (\ref{0020}), with some restrictions on $V(\rho)$, possess Galileo symmetry. For
    completeness, we list the manifest symmetries and the corresponding generators:

    \begin{itemize}
    \item Invariance under space and time translations,
       \begin{itemize}
       \item  Energy 
    \beq
    H = \int \rmdd r \Ee \ , \quad \Ee = \half \rho \boldnab \theta \cdot
    \boldnab \theta + V(\rho) 
    \label{0050}
    \eeq

    \item Momentum 
    \beq
    {\bf P} = \int \rmdd r \bcP \ , \quad
     \bcP =
    \rho
    \boldnab
    \theta = {\bf j}
    \label{0060}
    \eeq
       \end{itemize}

    \item Invariance under rotation

        \begin{itemize}
        \item Angular momentum 
    \beq
    J^{ij} = \int \rmdd r(r^i \Pp^j - r^j \Pp^i)
    \label{0055}
    \eeq
        \end{itemize}
    \end{itemize}

    \begin{itemize}
    \item  Galileo boost
           \bi 
           \item Boost generator
              \beq
              {\bf B} = t {\bf P} - \int \rmdd r \boldr \rho
              \label{0080}
              \eeq
           \ei
    \end{itemize}

    \begin{itemize}
    \item Invariance under global translations of the velocity potential, $\theta
    \rightarrow \theta + \alpha$, with $\alpha$ a constant.
         \bi
         \item Charge
    \begin{equation}
     M = \int \rmdd r \rho \label{0090}
    \end{equation}     \ei
    \end{itemize}
    Physically, the conservation of the charge  $M= \int d^3 x \, \rho$ (total mass)
    means that the center of mass of the fluid, ${\bf X} = \int d^3 x \, {\bf x} \rho/ M$
    moves with constant velocity,

    \begin{equation}
    \label{0100}
    M \frac{d\bf X}{dt} = \int d^3 x \, {\bf p},
    \end{equation}   
    where ${\bf p}$ denotes the total momentum of the fluid.

    Unexpectedly, extra symmetries were found in Refs.\cite{AJ,BJ,DB} and demonstrated
    that they are present on the model with a specific potential ($V(\rho)= g/\rho$) as
    well as for the free case. These extra symmetries are:

    \begin{itemize}
    \item Invariance under time rescaling $t \to e^{\omega} t$,
         \bi
         \item Time dilatation
    \begin{equation}
    D = tH - \int \rmdd r \rho \theta ,\label{0110}
    \end{equation}     \ei
    with the fields  transforming as

    \beqar
    \rho (t, \boldr) &\to& \rho_\omega (t, \boldr) = e^{-\omega} \rho
    (e^\omega t, \boldr)\nonumber\\
    \theta (t, \boldr) &\to& \theta_\omega (t, \boldr) =  e^{\omega} \theta
    (e^\omega t, \boldr). 
    \label{0120}
    \end{eqnarray}
    \end{itemize}

    \begin{itemize}
    \item Galileo antiboost,
         \bi
         \item Antiboost generator
    \beq
    {\vec G} = \int \rmdd r \left(\vec r \Ee - \half \rho \boldnab\theta^2
    \right)
    \eeq     \ei

    leading to 

    \begin{eqnarray}
    t&\to& T (t,\boldr)=t+\half {\pmb \omega} \cdot (\boldr +{\bf R} (t, \boldr))
    \nonumber
    \\
    \boldr &\to& {\bf R} (t, \boldr) =\boldr + {\pmb \omega} \theta(T, {\bf R}) 
    \label{eq:24}
    \end{eqnarray}
    where

    \begin{equation}
    \theta(T, {\bf R}) = \theta(t, \boldr - {\pmb \omega}t)+{\pmb \omega}.\boldr
    -\half {\pmb \omega}^2 t.
    \end{equation}
    \end{itemize}

    In Refs.\cite{BJ,DB} it was shown that only under the very specific density-dependent
    potential (\ref{0040}), the connection between the Galileo invariant system presented
    in this section, defined either in d=2 or d-space dimensions, and the relativistic
    membrane and its generalization to the d-brane  system in d=3 or (d+1)-space
    dimensions, appears . In view of this, it is remarkable to notice that the
    additional symmetries present on the Galileo invariant system in $d\geq1$ space
    dimensions with the interacting potential $V(\rho)= g/\rho$ are also present on the
    relativistic membrane and its generalization to the d-brane system in $d\geq2$ space
    dimensions.

    \section{Gauge Invariant Ideal Fluid Mechanics}

    To describe the gauge invariant hydrodynamics of an isentropic fluid, in d-space
    dimensions,
    we follow the presentation already discussed in the Introduction and the technique
    given in the appendix. Let us start with the Lagrangian given in (\ref{0020}) with an
    arbitrary potential $V(\rho)$. The canonical Hamiltonian is expressed as

    \be
    \label{MN50}
    {\cal H}_0 = \frac 12 \rho (\nabla\theta)^2 + V(\rho).
    \ee
    This is a (second-class) constrained theory but the primary constraints depend on our
    choice of the symplectic structure in (\ref{0020}), namely,

    \begin{mathletters}
    \ba
    \label{MN55A}
    {\cal L}^{(1)} &=& \int d^d r\;\left(\dot\rho \theta - \frac 12 \rho(\boldnab
    \theta)^2 - V(\rho)
    \right),\\
    \label{MN50B}
    {\cal L}^{(2)} &=& \int d^d r\;\left(- \rho \dot \theta - \frac 12 \rho(\boldnab
    \theta)^2 - V(\rho)
    \right),
    \ea
    \end{mathletters}
    which is deceptively trivial.
    Due to this difference, we have the following set of primary constraints,

    \begin{mathletters}
    \ba
    \label{MN60a}
    \varphi_\alpha^{(1)} &=&  \cases{\varphi_1^{(1)}&=$\;\;\pi_\rho - \theta$\cr
    \varphi_2^{(1)}&=$\;\;\pi_\theta$ \cr}\\
    \nonumber\\
    \label{MN60b}
    \varphi_\alpha^{(2)} &=&  \cases{\varphi_1^{(2)}&=$\;\; \pi_\theta + \rho$\cr
    \varphi_2^{(2)}&=$\;\; \pi_\rho $\cr}
    \ea
    \end{mathletters}
    where the superscript indices denote the distinct symplectic structures and the
    subscript indices denote the constraint family numbers. At this point, it is
    importante to notice that only one constraint, for each choice of sympletic
    structure $(i)$ in the set of constraints ($\varphi_\alpha^{(i)}$), can be lifted as a
    gauge symmetry generator. These constraints satisfy the following (second-class)
    Poisson algebra,

    \begin{mathletters}
    \ba
    \label{MN70a}
    \lbrace \varphi_1^{(1)}({\bf x}) , \varphi_2^{(1)}({\bf y}) \rbrace = -
    \delta^{(d)}({\bf x} - {\bf y}),\\
    \label{MN70b}
    \lbrace \varphi_1^{(2)}({\bf x}) , \varphi_2^{(2)}({\bf y}) \rbrace = +
    \delta^{(d)}({\bf x} - {\bf y}).
    \ea
    \end{mathletters}
    There are no more constraints in either cases. The Dirac Hamiltonian reads,

    \be
    \label{MN80}
    {\cal H}^{(i)}={\cal H}_0 +
    \lambda_1^{(i)}\varphi_1^{(i)}+\lambda_2^{(i)}\varphi_2^{(i)}\;\; ; \;\; i=1,2
    \;\;\mbox{(no sum)}
    \ee
    Temporal stability of the system demands,

    \be
    \label{MN90}
    \lbrace \varphi_\alpha^{(i)}({\bf x}) , {\cal H}^{(i)} \rbrace = 0 \;\;\; ;\;\;
    \alpha=1,2
    \ee
    leading to an explicit solution for the multipliers $\lambda_\alpha^{(i)}$ as,

    \begin{mathletters}
    \ba
    \label{MN100a}
    \lambda_\alpha^{(1)} = \cases{
    \lambda_1^{(1)}&=$\;\; \partial_i \left(\rho \partial_i\theta\right)$\cr
    \lambda_2^{(1)}&=$\;\; - \frac 12 (\nabla\theta)^2  - \frac{\partial
    V(\rho)}{\partial \rho}$\cr}\\
    \nonumber\\
    \label{MN100b}
    \lambda_\alpha^{(2)} = \cases{
    \lambda_1^{(2)}&=$\;\; - \frac 12 (\nabla\theta)^2  - \frac{\partial
    V(\rho)}{\partial \rho}$\cr
    \lambda_2^{(2)}&=$\;\; -\partial_i \left(\rho \partial_i\theta\right)$\cr}
    \ea
    \end{mathletters}
    Substitution of these results into the Dirac Hamiltonian leads to,

    \begin{mathletters}
    \ba
    \label{MN110a}
    {\cal H}^{(1)}&=&\frac 12 \rho (\nabla\theta)^2 + V(\rho)+
    \partial_i \left(\rho \partial_i\theta\right)( \pi_\rho - \theta) - \left(\frac 12
    (\nabla\theta)^2  + \frac{\partial V(\rho)}{\partial \rho}\right)\pi_\theta,\\
    \nonumber\\
    \label{MN110b}
    {\cal H}^{(2)}&=&\frac 12 \rho (\nabla\theta)^2 + V(\rho)-
    \partial_i \left(\rho \partial_i\theta\right) \pi_\rho  - \left(\frac 12
    (\nabla\theta)^2  + \frac{\partial V(\rho)}{\partial \rho}\right) \left(\pi_\theta +
    \rho\right).
    \ea
    \end{mathletters}
    As expected both descriptions differ by trivial terms. In spite of that the
    conversion procedure will lead to quite distinct systems with dramatic consequences,
    reflecting the interrelation between the different choices of symmetry generators for
    each case and the natural intrinsic symmetries of the potential.

    To disclose the hidden symmetry we follow the procedure reviewed in the appendix.  As
    mentioned after Eq.(\ref{total2}), one of the constraints is chosen as symmetry
    generator while the other is kept as gauge fixing condition.  For the systems above,
    (\ref{MN60a}) and (\ref{MN60b}), we will choose, without loss of physical contents,
    $\varphi_1^{(i)}({\bf x})$ to play the role of symmetry generator.  To choose the
    inhomogeneous constraints as symmetry generators is just a technical detail. The
    important consequence is that for the second case (i=2) the lift of the global
    translation invariance of the velocity potential to the local case is automatically
    provided by this choice of generator since 

\be
\delta \theta = \varepsilon\lbrace \theta, \varphi_1^{(2)}\rbrace = \varepsilon .
\ee
For the first sympletic structure, in opposition to the other case, this is not the natural choice, since $\varphi_1^{(1)}$ is the mass density translator not the potential translator,

\be
\delta \theta = \varepsilon\lbrace \theta, \varphi_1^{(1)}\rbrace = 0 .
\ee
We will however insist in follow this line of action to explore its consequences. As shown below this fact has striking effects over the
    first-class Hamiltonian structure.

    At this point all correction terms present on Eq.(\ref{HF}), given in appendix, can
    be explicitly computed just using the general relations

    \ba
    \label{MN120}
    A_1^{(i)} &=& \lbrace \varphi_1^{(i)}({\bf x}), H^{(i)}\rbrace,\nonumber\\
    A_2^{(i)} &=& \lbrace \varphi_1^{(i)}({\bf x}), A_1^{(i)}\rbrace,\nonumber\\
    A_3^{(i)} &=& \lbrace \varphi_1^{(i)}({\bf x}), A_2^{(i)}\rbrace,\nonumber\\
    \vdots \; &=& \; \vdots\nonumber\\
    A_n^{(i)} &=& \lbrace \varphi_1^{(i)}({\bf x}), A_{(n-1)}^{(i)}\rbrace\;\; ; \;\;
    n=1,2,\dots,
    \ea
    with $H^{(i)} = \int_y {\cal H}^{(i)}(y) dy$. The first sympletic structure ($i=1$)
    requires an infinite number of correction terms to lift the global translation
    symmetry, given by

    \ba
    \label{MN130}
    A_1^{(1)} &=& \lbrace \varphi_1^{(1)}({\bf x}), H^{(1)}\rbrace = \frac{\partial^2
    V}{\partial \rho^2}\pi_\theta,\nonumber\\
    A_2^{(1)} &=& \lbrace \varphi_1^{(1)}({\bf x}), A_1^{(1)}\rbrace = - \frac{\partial^3
    V}{\partial \rho^3}\pi_\theta - \frac{\partial^2 V}{\partial \rho^2},\nonumber\\
    A_3^{(1)} &=& \lbrace \varphi_1^{(1)}({\bf x}), A_2^{(1)}\rbrace = \frac{\partial^4
    V}{\partial \rho^4}\pi_\theta + 2\frac{\partial^3 V}{\partial \rho^3},\nonumber\\
    A_4^{(1)} &=& \lbrace \varphi_1^{(1)}({\bf x}), A_3^{(1)}\rbrace = - \frac{\partial^5
    V}{\partial \rho^5}\pi_\theta - 3\frac{\partial^4 V}{\partial \rho^4},\nonumber\\
    A_5^{(1)} &=& \lbrace \varphi_1^{(1)}({\bf x}), A_4^{(1)}\rbrace =  \frac{\partial^6
    V}{\partial \rho^6}\pi_\theta + 4\frac{\partial^5 V}{\partial \rho^5},\nonumber\\
    \vdots &=& \vdots\nonumber\\
    A_n^{(1)} &=& (-1)^{n+1}\left(\frac{\partial^{n+1}V}{\partial \rho^{n+1}}\pi_\theta +
    (n-1)\frac{\partial^{n+1} V}{\partial \rho^{n+1}}\right).
    \ea
    The second sympletic structure ($i=2$), on the other hand, only requires two
    correction terms to covariantize the canonical structure,

    \ba
    \label{MN140}
    A_1^{(2)} &=& \lbrace \varphi_1^{(2)}({\bf x}), H^{(2)}\rbrace = -\partial \int_y
    \partial[\rho\delta^{(d)}({\bf x} - {\bf y})]\pi_\rho,\nonumber\\
    A_2^{(2)} &=& \lbrace \varphi_1^{(2)}({\bf x}), A_1^{(2)}\rbrace = -\partial \int_y
    \partial[\rho\delta^{(d)}({\bf x} - {\bf y})],\\
    A_3^{(2)} &=& \lbrace \varphi_1^{(2)}({\bf x}), A_2^{(2)}\rbrace =0.\nonumber
    \ea
    As discussed above, the minimal correction for the second Hamiltonian structure to produce a
    gauge
    invariant formulation happens because the constraint ($\varphi_1^{(2)}$) promoted to
    first-class lifts the global
    translation symmetry as a local gauge symmetry. This is not true for the other
    sympletic structure.  Bringing these results into the projected Hamiltonian derived
    in the appendix (Eq.(\ref{HF})), the gauge invariant Hamiltonians correspondent to
    those different symplectic structures are,

    \begin{mathletters}
    \ba
    \label{MN150a}
    \tilde{\cal H}^{(1)}&=&\frac 12 \rho (\nabla\theta)^2 + V(\rho)+
    \partial_i \left(\rho \partial_i\theta\right)( \pi_\rho - \theta) - \frac 12
    (\nabla\theta)^2\pi_\theta  + \sum_{n=1}^\infty \frac{(-1)^n}{n!}\frac{\partial^n
    V}{\partial \rho^n} \pi_\theta^n \\
    \nonumber\\
    \label{MN150b}
    \tilde{\cal H}^{(2)}&=&\frac 12 \rho (\nabla\theta)^2 + V(\rho)-
    \partial_i \left(\rho \partial_i\theta\right) \pi_\rho  - \left(\frac 12
    (\nabla\theta)^2  + \frac{\partial V(\rho)}{\partial \rho}\right) \left(\pi_\theta +
    \rho\right) \nonumber\\
    &+& \pi_\rho\partial_{\bf x}\int_y \partial_{\bf y}[\rho\delta^{(d)}({\bf x} - {\bf
    y})]\pi_\rho
    - \frac 12 \pi_\rho^2\partial_x\int_y \partial_{\bf y}[\rho\delta^{(d)}({\bf x} -{\bf
    y})].
    \ea
    \end{mathletters}
    It is noticiable that the second symplectic structure indeed contains a finite number of corrections while the first one  contains infinite correction terms which, however, can be summed up to a closed form, as shown below.

    The infinitesimal gauge transformations associated to each of these invariant models
    are now computed,

    \ba
    \label{MN160}
    \delta^{(i)} \rho &=&\, \varepsilon\lbrace \rho, \varphi_1^{(i)}\rbrace
    =\cases{\delta^{(1)} \rho &=$\varepsilon$\cr
    \delta^{(2)} \rho &= 0\cr}\\
    \nonumber\\
\label{MN161}
    \delta^{(i)} \theta &=&\, \varepsilon\lbrace \theta, \varphi_1^{(i)}\rbrace
    =\cases{\delta^{(1)} \theta &=$0$\cr
    \delta^{(2)} \theta &= $\varepsilon$\cr}\\
    \nonumber\\
    \delta^{(i)} \pi_\rho &=& \varepsilon\lbrace \pi_\rho, \varphi_1^{(i)}\rbrace
    =\cases{\delta^{(1)} \pi_\rho &=0\cr
    \delta^{(2)} \pi_\rho&=$-\varepsilon$\cr}\\
    \nonumber\\
    \delta^{(i)} \pi_\theta &=& \varepsilon\lbrace \pi_\theta, \varphi_1^{(i)}\rbrace
    =\cases{\delta^{(1)} \pi_\theta &=$\varepsilon$\cr
    \delta^{(2)} \pi_\theta&=0\cr}
    \ea
    As mentioned, the local translation of the velocity potential $\theta$ is directly
    realized with the choice of the second sympletic structure, Eq. (\ref{MN161}), while
    the choice of the first structure, produces a local shift on the mass density, Eq.(\ref{MN160}).

    It is important to observe that these reformulations of fluid dynamics as gauge
    theories were obtained here without especifying the dependence of the interacting
    potential on $\rho$. Of course this is not a remarkable result for the question put
    forward by the constraint conversion mechanism, but it might be helpful to shed light
    on the question related to the connection of the Galileo invariant system with the
    relativistic membranes and its generalizations to the d-branes system, which only
    appears under the specific potential (\ref{0040}).

    It is interesting at this juncture, in order to put our work in perspective, to
    relate our results with the gauge formulation obtained in Ref.\cite{NF} through the
    enlargement of the phase space with Wess-Zumino (WZ) variables in the context of BFFT
    formalism \cite{BFFT,IJMP}. It is worth mentioning that in the BFFT formalism, unlike \cite{Mitra,VT},
all constraints are covariantized and it is not clear to us at this point how this formalism will produce the distinctic results showed above.  In \cite{NF}, the authors proposed an invariant
    Lagrangian, given in terms of an infinite number of corrections, that can be written
    in closed form as,

    \be
    \label{MN170}
    {\cal L} = \theta\dot\rho + \varphi\dot\theta - \frac 12 (\rho -
    \varphi)(\nabla\theta)^2 - V(\rho-\varphi),
    \ee
    where $\varphi$ is the WZ variable. The corresponding first-class Hamiltonian is
    given by

    \be
    \label{MN180}
    {\cal H} = \frac 12 (\rho - \varphi)(\nabla\theta)^2 + V(\rho-\varphi)
    \ee
    while the full set of constraints for this theory is given by

    \ba
    \label{MN190}
    \phi_1 &=& \theta - \pi_\rho,\nonumber\\
    \phi_2 &=& \pi_\varphi,\\
    \phi_3 &=& \pi_\theta - \varphi.\nonumber
    \ea
    Although the Dirac matrix is singular, some constraints have nonvanishing Poisson
    brackets, This is so because the authors of \cite{NF} failed to recognize $ \phi_2$ as one of the primary constraints of the theory. Due to this, the set of constraints (\ref{MN190}) can be splited in its
    first and second-class components. Redefining the constraints as,

    \ba
    \label{MN200}
    \tilde\phi_1 &=& \theta - \pi_\rho - \pi_\varphi,\nonumber\\
    \tilde\phi_2 &=& \pi_\varphi,\\
    \tilde\phi_3 &=& \pi_\theta - \varphi,\nonumber
    \ea
    turns the $\tilde\phi_1$ into first-class while $\tilde\phi_2$ and $\tilde\phi_3$
    remain second-class constraints satisfying the canonical Poisson bracket,

    \be
    \label{MN21}
    \lbrace\tilde\phi_2, \tilde\phi_3\rbrace = \delta^{(d)}({\bf x} -{\bf y}).
    \ee
    Taking these constraints equal to zero in a strong way, the Dirac brackets among the
    phase space variables can be computed. Due to the Maskawa-Nakajima theorem\cite{MN},
    the Dirac brackets are bound to be canonical. In view of this, the gauge invariant Hamiltonian
    becomes

    \be
    \label{MN220}
    {\cal H} = \frac 12 (\rho - \pi_\theta)(\nabla\theta)^2 + V(\rho-\pi_\theta),
    \ee
    and the remaining (first-class) constraint, namely,

    \be
    \label{MN230}
    \tilde\varphi_1= \theta - \pi_\rho,
    \ee
    becomes the generator of symmetry. This system is equivalent to the solution obtained
    by us with the choice of the first-symplectic struture following the gauge unfixing
    scheme\cite{Mitra,VT} after summing up the infinite terms of (\ref{MN150a}). 

    \section{Conclusion}

The description of a relativistic membrane in terms of a reduced
    fluid dynamics\cite{BH,AJ}, has intensified the study of this topic over the last
    years, establishing the connection between this Galileo
    invariant system in d-space dimensions and relativistic membranes and its
    generalizatons to the d-branes system in (d+1)-space dimensions. This led to the identification of two extra
    symmetries but the resulting hidrodynamical system lacks gauge symmetry. Inspired by this result, we argue
    that the symmetry present on the gauge description of relativistic membrane may be recovered on theory of isentropic fluid dynamics, after the
    reduction process\cite{BH,AJ}. In this paper the gauge invariant version for the
    fluid theory was obtained and the gauge symmetry recovered by using the gauge
    unfixing technique\cite{Mitra,VT}.

The use of this technique has made it clear that the final result is dramatically dependent on the initial choice of the symplectic structure.  This is so because, in the present case the symmetry restored is that of translation of the velocity potential.  A quite simple procedure is obtained using the constraint that effectively translate that variable (the second symplectic structure in this case) leading to just two correction terms in the covariantization process.  This seems natural since this is the constraint that elevates the global symmetry originally defined in the system, i.e., the translation of the velocity potential, Eq.(\ref{0090}). On the other hand, when the first sympletic structure is adopted, the gauge generator involved is not the velocity translator but the mass density translator instead.  Since the potential appearing in the original Hamiltonian (\ref{MN50})is only dependent on the density $\rho$, each application of this generator produces a derivative of this term. Consequently, in order to perform the task of lifting the global symmetry into a local symmetry, an infinite number of terms is required.

A further important point emphasized in this paper is that this embedded
    symmetry does not lie on the WZ sector, as proposed by phase-space enlarging techniques, but it lies on the
    original phase space. That makes it clear that the gauge symmetry manifest in the
    relativistic theory of membranes correspond to the local translations symmetry of the velocity potential in the fluid dynamics model, as explained in Section III. With this result we
    established a complete connection between Galileo invariant system and d-branes system.

\noindent ACKNOWLEDGMENTS: This work is partially supported by CNPq, CAPES, FAPERJ, FAPEMIG and
FUJB, Brazilian Research Agencies.

    \appendix

    \section{The gauge unfixing formalism}

    In the Hamiltonian reduction of the second-class constrained systems,
    the constraints are classified as primary and secondary\cite{PD}. 
    Secondary constraints are consistency conditions following from
    the primary constraints. This consistency algorithm must be implemented
    until all independent constraints are obtained. Consider a Hamiltonian
    system in a 2M dimensional phase space $(q,p)$ with an even number of
    second-class constraints $\phi_\alpha\approx 0 $ ($\alpha=1,2,\dots,2N$, where $N <
    M$)
    described as,

    \be
    \label{A00000}
    H=H_c+u_\alpha\phi_\alpha, \,\,\,\, \alpha=1,2,\dots, 2N,
    \ee
    where $H_c$ represents the canonical Hamiltonian and $u_\alpha$ are the Lagrange
    multipliers.
    For second-class constraints $\phi_\alpha$, the $u_\alpha$ can be determined
    everywhere by demanding,

    \ba
    \label{A00010}
    \lbrace\phi_\beta , H\rbrace &=& \lbrace\phi_\beta , H_c\rbrace +
    u_\alpha\lbrace\phi_\beta ,\phi_\alpha\rbrace\approx 0,\nonumber\\
    u_\alpha &\approx& - \frac{\lbrace\phi_\beta , H_c\rbrace}{\lbrace\phi_\beta
    ,\phi_\alpha\rbrace},
    \ea
    for $\alpha=1,2,\dots,2N$. Using these results into the total Hamiltonian
    (\ref{A00000}), we then have the consistency condition for $\phi_\alpha$,

    \be
    \label{A00020}
    \lbrace\phi_\alpha, H\rbrace = G_{\alpha\beta}\phi_\beta,
    \ee
    where $G_{\alpha\beta}$ are structure constants. This completes Dirac's consistency
    algorithm.

    The main idea of the gauge unfixing procedure is to consider half of the second-class
    constraints as gauge fixing conditions\cite{Mitra,VT,HT} over the remaining first-class,
    gauge generators, constraints. 
    Next the first-class Hamiltonian is obtained in a systematic way \cite{VT} using a properly constructed projector operator.  Let us consider a
    system with two second-class constraints, $\phi_1$ and  $\phi_2$, satisfying the
    following Poisson brackets algebra,

    \be
    \label{A00030}
    C = \{ \phi_1, \phi_2 \}.
    \ee
    Redefining the constraints as

    \ba
    \label{A00040}
    \xi \equiv C^{-1} \phi_1,\nonumber\\
    \psi\equiv \phi_2,
    \ea
    we have

    \be
    \label{A00050}
    \{\xi, \psi\}= 1 + \lbrace C^{-1},\psi\rbrace C\xi,
    \ee
    so that $\xi$ and $\psi$ are canonically conjugate on the surface defined by $\xi=0$.
    The total Hamiltonian, following (\ref{A00000}) is

    \begin{eqnarray}   
    \label{total2}
    H=H_c+u_1\xi+u_2\psi.
    \end{eqnarray}
    Let us maintain only $\xi$ as a natural constraint relation. This will allow us to
    obtain
    the first-class system relative to it.  However notice that at first, $\{\xi,H\} \neq
    0$,
    so that in principle, $\xi$ and $H$,  do not satisfy a first-class algebra. The
    proper
    first-class Hamiltonian can be expressed by the Poisson projection \cite{VT}

    \ba
    \label{HF}
    \tilde{H} &=& H - \psi \{ \xi, H \} + {1\over 2!} \psi^2 \{ \xi ,  \{ \xi, H \} \}
    -  {1\over 3!} \psi^3 \{\xi, \{ \xi , \{ \xi, H \} \} + \dots \nonumber\\
    &=& : \exp^{-\psi \xi} : H,
    \ea
    which satisfies the first-class condition

    \begin{eqnarray}
    \label{fc}
    \{\xi,\tilde{H}\} =0.
    \end{eqnarray}
    The first-class Hamiltonian $\tilde{H}$ was rewritten in terms of a projection form
    with $\psi$ respecting the ordering rule defined by Poisson bracket operation above. 

    As discussed, this formalism converts a second-class system into first-class directly
    into the original phase-space. This result is equivalent to a partial gauge-fixing
    of the WZ sector when the converting formalism using the phase-space enlargement Faddeev-Shatshvili idea is adopted \cite{CNCW}. In this context, the ambiguity in the
    choice of the first-class constraint is related to different choices of fixing
    the WZ sector.

    \end{document}